\documentclass{jps-cp}
\usepackage{txfonts}

\title{\textit{Ab Initio} Calculations of $^6$He + $p$ Reactions}

\author{Jakub \textsc{Herko}$^{1}$, Petr \textsc{Navr\'atil}$^{1,2}$, Konstantinos \textsc{Kravvaris}$^{3}$, Sofia \textsc{Quaglioni}$^{3}$, and Guillaume \textsc{Hupin}$^{4}$}

\inst{$^{1}$TRIUMF, 4004 Wesbrook Mall, Vancouver, British Columbia V6T 2A3, Canada \\
$^{2}$University of Victoria, 3800 Finnerty Road, Victoria, British Columbia V8P 5C2, Canada \\
$^{3}$Lawrence Livermore National Laboratory, 7000 East Ave, Livermore, California 94550, USA \\
$^{4}$Universit\'e Paris-Saclay, CNRS/IN2P3, IJCLab, 91405 Orsay, France}

\email{jherko@triumf.ca}

\abst{The no-core shell model with continuum (NCSMC) is an \textit{ab initio} framework for description of nuclear structure and reactions. Recent developments of the NCSMC allow for calculations of charge-exchange and nucleon-transfer reactions. We have applied the NCSMC to calculate cross sections of the ($p,p$), ($p,n$), and ($p,t$) reactions on radioactive $^6$He, using chiral nuclear forces as input.}

\kword{\textit{ab initio} no-core shell model with continuum, $^6$He + $p$ cross sections}

\begin{document}
\maketitle

\section{Introduction}

The no-core shell model with continuum (NCSMC) \cite{baroni2013prl,baroni2013prc} is an \textit{ab initio} framework for a unified description of nuclear structure and reactions. It extends the no-core shell model (NCSM) \cite{navratil2000} by augmenting the model space with states in which the nucleons are distributed between clusters. Recent computational developments \cite{kravvaris2024} enable inclusion of different distributions of the nucleons between the clusters (mass/charge partitions or, simply, partitions) in a single coupled-channels calculation. This allows for calculations of charge-exchange and nucleon-transfer reactions. We perform the first NCSMC calculation coupling three partitions, namely $^6$He + $p$, $^6$Li + $n$, and $^4$He + $^3$H, using chiral nuclear forces as input \cite{herko2026}. This enables description of the ($n,n$), ($n,p$), and ($n,t$) reactions on $^6$Li and ($p,p$), ($p,n$), and ($p,t$) reactions on $^6$He. Here we present the cross sections of the proton-induced reactions on $^6$He.

\section{Methods}

In this work the many-nucleon system is described by an intrinsic Hamiltonian $H$ consisting of the intrinsic kinetic energy and the chiral nucleon-nucleon interaction at next-to-next-to-next-to-leading order of Ref. \cite{entem2003}. We soften the interaction using a similarity-renormalization-group transformation \cite{bogner2007} with flow parameter $\lambda_{\rm SRG}=2.15$ $\textrm{fm}^{-1}$, neglecting induced three- and higher-body forces.

To obtain many-body static solutions for this Hamiltonian, we employ the NCSM, where the Schr\"odinger equation $H|\Psi\rangle=E|\Psi\rangle$ is solved as an eigenvalue problem for the Hamiltonian represented as a matrix in a harmonic oscillator (HO) many-body basis. The basis is characterized by HO frequency $\Omega$, and its size is determined by the maximum number of excitation quanta, $N_{\rm max}$.

To account for continuum degrees of freedom, we employ the NCSMC, where the wave function of the many-body system is expanded in terms of NCSM eigenstates of $H$ and continuous microscopic binary-cluster states within the resonating-group method (RGM) \cite{tang1978}. In these RGM-type states, the intrinsic wave functions of the clusters are also eigenstates of $H$ calculated in the NCSM framework \cite{quaglioni2008}. The discrete coefficients and continuous amplitudes of the expansion are obtained by solving the NCSMC equations \cite{baroni2013prl,baroni2013prc}. These equations can be solved for either bound or scattering states by adopting the appropriate asymptotic form of the continuous expansion amplitudes, which represent the relative motion of the clusters. For scattering states, this asymptotic form determines the scattering matrix, from which cross sections can be obtained.

\section{Results}

In the present application, the aggregate system is $^7$Li and we consider three partitions: $^4$He + $^3$H, $^6$Li + $n$, and $^6$He + $p$. In the NCSM calculations, we used HO model spaces up to $N_{\rm max}=12$ for $^4$He and $^3$H, up to $N_{\rm max}=10$ for $^6$Li, $^6$He, and the negative-parity states of $^7$Li, and up to $N_{\rm max}=11$ for the positive-parity states of $^7$Li. We adopt $\hbar\Omega=20$ MeV. We include the lowest twelve negative-parity and the lowest six positive-parity NCSM eigenstates of $^7$Li. For $^4$He and $^3$H, we include only their ground states. For $^6$Li, we include four states with angular momenta, parities, and isospins $J_1^{\pi_1}T_1=1^+0,3^+0,0^+1,2^+1$, while for $^6$He we include two states with $J_1^{\pi_1}T_1=0^+1,2^+1$.

\subsection{Phenomenological adjustments}

\begin{figure}[tbh]
\centering
\includegraphics[width=0.9\linewidth]{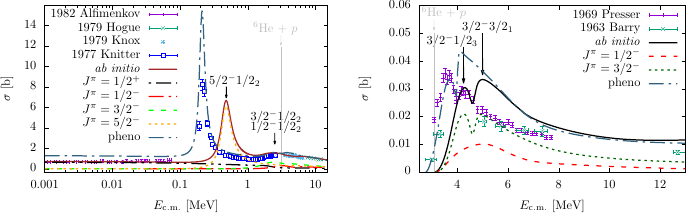}
\caption{The cross sections of the elastic scattering of neutrons on $^6$Li (left) and the $^6$Li($n,p)^6$He reaction (right) as functions of the kinetic energy in the center-of-mass frame calculated within the NCSMC (solid line) along with experimental data \cite{alfimenkov1982,hogue1979,knox1979,knitter1977,presser1969,barry1963}, results calculated by restricting the total angular momentum and parity $J^{\pi}$ to specific values (dashed lines), and results calculated phenomenologically adjusting NCSM energies (denoted by ``pheno"). Resonances corresponding to the peaks in the calculated cross sections are indicated. The vertical gray line denotes the calculated $^6$He + $p$ threshold.}
\label{f1}
\end{figure}

The left panel of Fig. \ref{f1} shows the calculated cross section of the elastic scattering $^6$Li($n,n)^6$Li as a function of the kinetic energy in the center-of-mass frame $E_{\rm c.m.}$ (solid line) along with experimental data. There are two visible peaks corresponding to the $5/2^-1/2_2$ and $3/2^-1/2_2$ resonances in $^7$Li with a small contribution from the $1/2^-1/2_2$ resonance. The two-peak structure is reflected in the experimental data, albeit at slightly different energies.

The right panel of Fig. \ref{f1} shows the cross section of the reaction $^6$Li($n,p)^6$He (going from the ground state of $^6$Li to the ground state of $^6$He) along with experimental data. The calculated cross section has two peaks arising from $3/2^-1/2$ and $3/2^-3/2$ resonances with a contribution from a $1/2^-1/2$ resonance. The peaks are mirrored in the Presser data \cite{presser1969}, however, again, at lower energy.

The NCSMC equations are expressed via integral kernels, namely the norm and Hamiltonian kernels \cite{baroni2013prc}, with the Hamiltonian kernel being a function of the NCSM energies of the target and the aggregate nucleus. To better describe the ($n,n$) and ($n,p$) cross sections, we consider phenomenological adjustments of these energies.

First we adjusted the NCSM energies of $^6$Li and $^6$He to reproduce the experimentally observed spectra and the $^6$Li + $n$ and $^6$He + $p$ decay thresholds relative to the $^4$He + $^3$H threshold. Next, we adjusted the energies of the states $5/2^-1/2_2$ and $3/2^-3/2_1$ in $^7$Li to reproduce their distances to the $^6$Li + $n$ threshold. In this way we reproduced the positions of the peak in the ($n,n$) cross section and the second peak in the ($n,p$) cross section. We also adjusted the energy of a predicted state $3/2^-1/2_3$ to reproduce the position of the first peak in the ($n,p$) cross section and the energies of the states $3/2^-1/2_2$ and $1/2^-1/2_2$ to improve the agreement with the experimental ($n,n$) cross section at high energy and of a predicted state $1/2^-1/2_3$ to better describe the overall shape of the ($n,p$) cross section.

The resulting cross sections are shown in Fig. \ref{f1} and denoted by ``pheno”. The new ($n,n$) cross section presents a peak at the same energy as the experimental cross section and agrees fairly well with the high-energy data, while the new ($n,p$) cross section reproduces the position of the two peaks.

\subsection{Cross sections of $^6$He + $p$ reactions}

\begin{figure}[tbh]
\centering
\includegraphics[width=0.9\linewidth]{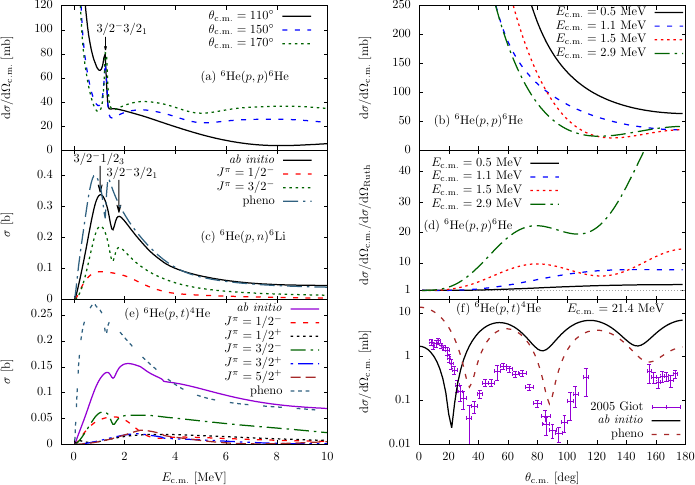}
\caption{(a) Differential cross sections of the elastic scattering of protons on $^6$He as functions of the kinetic energy in the center-of-mass frame $E_{\rm c.m.}$ for different scattering angles $\theta_{\rm c.m.}$ calculated phenomenologically adjusting NCSM energies. The peak corresponds to the $3/2^-3/2_1$ resonance. (b) Differential cross sections of the elastic scattering of protons on $^6$He as functions of $\theta_{\rm c.m.}$ calculated phenomenologically adjusting NCSM energies. (c) Calculated cross section of the $^6$He($p,n)^6$Li reaction (denoted by ``\textit{ab initio}") along with results calculated by restricting $J^{\pi}$ to specific values (dashed lines) and result calculated phenomenologically adjusting NCSM energies (denoted by ``pheno”). Resonances corresponding to the peaks are indicated. (d) Ratios of the differential cross sections of the elastic scattering of protons on $^6$He to the Rutherford cross section. (e) Cross section of the $^6$He($p,t)^4$He reaction. (f) Differential cross section of the $^6$He($p,t)^4$He reaction along with experimental data \cite{giot2005}.}
\label{f2}
\end{figure}

To provide theoretical predictions for a comparison with future measurement of $^6$He + $p$ reactions planned at TRIUMF, Fig. \ref{f2} shows calculated cross sections of such reactions. Panels (a), (b), and (d) show differential cross sections of the elastic scattering of protons on $^6$He as functions of $E_{\rm c.m.}$ for different scattering angles $\theta_{\rm c.m.}$ [panel (a)] and as functions of $\theta_{\rm c.m.}$ for different values of $E_{\rm c.m.}$ [panel (b)] along with ratios to the Rutherford cross section [panel (d)], calculated with the phenomenological adjustments. For low $E_{\rm c.m.}$ ($\theta_{\rm c.m.}$) the cross sections increase with decreasing $E_{\rm c.m.}$ ($\theta_{\rm c.m.}$) in accordance with the Rutherford formula, and they have a peak corresponding to the $3/2^-3/2_1$ resonance [panel (a)]. For the lowest energy the calculated cross section is close to the Rutherford cross section [the ratio in panel (d) is close to 1], and with increasing $E_{\rm c.m.}$ and $\theta_{\rm c.m.}$ the discrepancy from the Rutherford cross section increases.

Panels (c) and (e) of Fig. \ref{f2} show cross sections of the reactions $^6$He($p,n)^6$Li and $^6$He($p,t)^4$He calculated without (denoted by ``\textit{ab initio}") and with (denoted by ``pheno") the phenomenological adjustments. The cross sections have two peaks corresponding to the $3/2^-1/2_3$ and $3/2^-3/2_1$ resonances with contributions from other $J^{\pi}$ channels, as indicated by results calculated by restricting $J^{\pi}$ to specific values (dashed lines). After the phenomenological adjustments, the overall shape of the ($p,n$) cross section changed only slightly, and the ($p,t$) cross section is enhanced at low energy preserving the two-peak shape.

Panel (f) of Fig. \ref{f2} shows the differential cross section of the $^6$He($p,t)^4$He reaction as a function of $\theta_{\rm c.m.}$ for $E_{\rm c.m.}=21.4$ MeV calculated without (solid line) and with (dashed line) the phenomenological adjustments along with experimental data. Without the phenomenological adjustments the calculation reproduces the overall shape of the measured cross section, however, it does not reproduce the positions of the minima and maxima. With the phenomenological adjustments it better reproduces the positions of the minima and maxima. It overestimates the experimental values, but the energy $E_{\rm c.m.}=21.4$ MeV is beyond applicability of the current setup due to omitted channels which are open at this high energy. However, there are no experimental data at lower energy for comparison.

\section{Conclusion}

This work presents the first NCSMC coupled-channels calculation that includes three partitions, in particular $^6$He + $p$, $^6$Li + $n$, and $^4$He + $^3$H. The coupling of the partitions allowed us to calculate cross sections of the reactions $^6$He($p,n)^6$Li and $^6$He($p,t)^4$He. We also calculated differential cross sections of the elastic scattering of protons on $^6$He. The cross sections are to be compared with future results of an experiment planned at TRIUMF to probe resonances in $^7$Li above the $^6$He + $p$ threshold by impinging a beam of radioactive $^6$He onto a target with hydrogen thereby studying $^6$He + $p$ reactions in inverse kinematics.

\end{document}